\documentclass{ws-p8-50x6-00}

\usepackage{amssymb}
\usepackage{axodraw}

\begin{document}

\title{Fermion transmutation---a renormalization effect in gauge theory}

\author{TSOU Sheung Tsun}

\address{Mathematical Institute, University of Oxford,\\
  24-29 St. Giles', Oxford OX1 3LB, United Kingdom\\E-mail: 
  tsou\,@\,maths.ox.ac.uk}

\author{CHAN Hong-Mo}

\address{Rutherford Appleton Laboratory,\\
  Chilton, Didcot, Oxon, OX11 0QX, United Kingdom\\
E-mail: chanhm\,@\,v2.rl.ac.uk}  


\maketitle

\abstracts{
A new category of phenomena is predicted in which fermions of different
flavours can transmute into one another, for example $e \to \mu$ 
or $e \to \tau$, 
as a consequence of the `rotating' mass matrix due to renormalization. As
examples, calculations will be presented for various such processes.  Some
of these could be accessible to experiments in the near future.}

\section{Introduction}
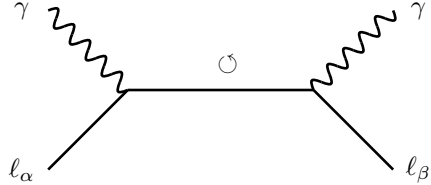
\begin{figure}
\vspace*{-8mm}
\begin{center}
{\unitlength=1.0 pt \SetScale{1.0} \SetWidth{1.0}
\begin{picture}(150,100)(0,0) 
\Line(30,50)(100,50)
\Photon(30,50)(0,80){3}{5}
\Photon(100,50)(130,80){3}{5}
\Line(0,20)(30,50)
\Line(100,50)(130,20)

\Text(-10,20)[]{$\ell_\alpha$}
\Text(140,20)[]{$\ell_\beta$}
\Text(-10,80)[]{$\gamma$}
\Text(140,80)[]{$\gamma$}
\Text(68,60)[]{$\circlearrowleft$}

\end{picture} }
\end{center}
\vspace*{-6mm}
\caption{Photo-transmutation of leptons.}
\label{compton}
\vspace*{-6mm}
\end{figure}

By `fermion transmutation' we mean a process in which a fermion changes
its generation index
as a direct consequence of the rotation of the mass matrix,
and not as a secondary effect such as $e-\mu$ conversion via FCNC.
Examples of transmutation are: $e \to \mu,\ e \to \tau,\ \mu \to
\tau$, which can occur e.g.\ in a Compton-like process schematically
represented in Figure \ref{compton}.

That the fermion mass matrix rotates, by which we mean that it
undergoes unitary transformations as the scale changes, can be seeen
from the renormalization group equation, both in the standard model
(SM) and in the dualized standard model (DSM).

An earlier talk\cite{hm} summarizes our
earlier work on DSM\cite{dsm}.
In this talk I shall report on two effects, namely transmutational
decays\cite{impromat} and photo-transmutations\cite{photrans}, 
in both SM and DSM, but with emphasis on DSM.

\section{Mass matrix rotation}
The SM renormalization group equation\cite{rge} for the charged lepton
mass matrix $L$ has a term which, given that the leptonic MNS mixing 
matrix\cite{mns} $U$ is nontrivial\cite{expmix}, rotates it as the
scale $\mu$ changes.  The linearized
equation:
\begin{equation}
\frac{dL}{d\mu} = \frac{3}{128 \pi^2} \frac{1}{246^2} (ULU^\dag)
(ULU^\dag)^\dag L + \cdots,
\label{smrge}
\end{equation}
where $ULU^\dag=N$ the neutrino (Dirac) mass matrix, already shows that
even if $L$ is diagonal at a chosen scale it
cannot remain so at all other scales.  The magnitude of the off-diagonal
elements will depend on poorly known or unknown quantities such as the
mixing $U$ and the Dirac mass $m_3$ of the heaviest neutrino.  If we
take the present popular theoretical biases of $U$ 
bimaximal and $m_3 \sim m_t$,
then (\ref{smrge}) gives for each decade change in energy:
\begin{eqnarray*}
\langle \mu | \tau \rangle & {\rm changes\ by} & \sim 5.5 \times
10^{-3}\ {\rm GeV} \\
\langle e | \tau \rangle & {\rm changes\ by} & \sim 1.8 \times 
10^{-7}\ {\rm GeV}\\
\langle e | \mu \rangle & {\rm changes\ by} & \sim 1.1 \times 
10^{-8}\ {\rm GeV}
\end{eqnarray*}

\begin{figure}
\centering
\epsfxsize120pt
\includegraphics[angle=-90,scale=0.6]{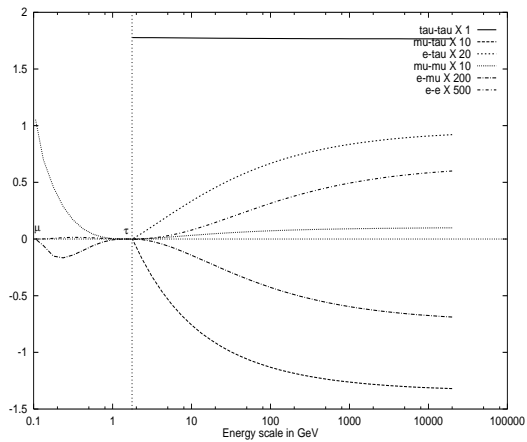}
\caption{Rotating mass matrix elements in GeV for charged 
leptons in the DSM scheme.}
\label{masmat}
\vspace*{-5mm}
\end{figure}

In the DSM\cite{dsm}, the fermion mass matrix is of the following
factorized form: 
\begin{equation}
m = m_T \left( \begin{array}{c} x \\ y \\ z \end{array} \right)
      (x, y, z),
\label{mmat}
\end{equation}
where $m_T$ is essentially the mass of the heaviest generation.  Under
renormalization $m$ remains factorized, but the vector $(x,y,z)$
changes as
\begin{equation}
\frac{d}{d\mu} \left( \begin{array}{c} x \\ y \\ z \end{array} \right)
   = \frac{5}{32 \pi^2} \rho^2 \left( \begin{array}{c} x_1 \\ y_1 \\ z_1
      \end{array} \right),
\label{dsmrge}
\end{equation}
where $\rho$ is a (fitted) constant and
\begin{equation}
x_1 = \frac{x(x^2-y^2)}{x^2+y^2} + \frac{x(x^2-z^2)}{x^2+z^2},
   \ \ \ {\rm cyclic}.
\end{equation}
The off-diagonal elements have been calculated explicitly, using 3
free parameters determined by fitting experimental mass and mixing
parameters (giving sensible predictions for the remaining
paramenters)\cite{phenodsm}. These are shown in Figure \ref{masmat}.  
Hence the
results we report below are entirely parameter-free.

\section{Definition of lepton states}
To define lepton states one must diagonalize the mass matrix, but
since the eigenvectors depend on scale, there is no canonical recipe
for doing so.

We suggest two quite different schemes for exploration.  In the fixed
scale diagonalization scheme (FSD) the state vectors are fixed at a
chosen scale.  As a result, the mass matrix is diagonal only at that
scale.  Such a scheme is applicable to SM, where it is tacitly assumed
in most calculations.

For the DSM we use a more sophisticated scheme which we may call
step-by-step diagonalization (SSD)\cite{dsm,impromat,photrans}.  
It is a working criterion in 
which the lepton state vectors are always orthogonal and the mixing
matrix is always unitary.   There are 3 main steps:\\
\hspace*{\parindent}$\bullet$ run the $3 \times 3$ mass matrix to a
scale equal to mass of heaviest generation;\\
\hspace*{\parindent}$\bullet$ corresponding eigenvector is its state vector;\\
\hspace*{\parindent}$\bullet$ repeat with the remaining $2 \times 2$ 
submatrix.\\
In contrast to FSD, the charged
lepton mass matrix here is diagonal at scales of $m_\tau,m_\mu,m_e$,
as seen in Figure \ref{masmat}.

\section{Transmutational decays}
With the lepton states defined as in \S 3 and the 
rotations obtained in \S2 we can now study transmutational processes,
the most obvious of which are decays.

By expanding the fermion
propagator we get for $\ell_\alpha \to \ell_\beta,\ \alpha \ne \beta$:
\begin{equation}
{\rm transmutation}/{\rm diagonal} \sim {\langle \alpha |
\beta \rangle}/{E},
\end{equation}
where $E$ is a typical energy for the decay.  Estimates are in 
Table \ref{decays}.
\begin{table}[t]
\caption{Branching ratios of transmuational decays.}\label{decays}
\begin{center}
\footnotesize
\begin{tabular}{|l|c|c|c|}
\hline 
 
\raisebox{0pt}[13pt][7pt]{Decays} & 
\raisebox{0pt}[13pt][7pt]{SM est.} & 
\raisebox{0pt}[13pt][7pt]{DSM est.} & 
 
\raisebox{0pt}[13pt][7pt]{Expt limit} \\
 
\hline

\raisebox{0pt}[13pt][7pt]{$Z^0 \to \tau^- \mu^+$} & 
 \raisebox{0pt}[13pt][7pt]{$4 \times 10^{-10}$} &
\raisebox{0pt}[13pt][7pt]{$4 \times 10^{-8}$} & 
 
\raisebox{0pt}[13pt][7pt]{$1.2 \times 10^{-5}$} \\

\raisebox{0pt}[13pt][7pt]{$\pi^0 \to \mu^- e^+$} & 
\raisebox{0pt}[13pt][7pt]{negligible} &
\raisebox{0pt}[13pt][7pt]{$3 \times 10^{-9}$} & 
 
\raisebox{0pt}[13pt][7pt]{$1.7 \times 10^{-8}$} \\

\raisebox{0pt}[13pt][7pt]{$\psi \to \mu^+ \tau^-$} & 
\raisebox{0pt}[13pt][7pt]{$1 \times 10^{-8}$} & 
\raisebox{0pt}[13pt][7pt]{$6 \times 10^{-6}$} & 
 
\raisebox{0pt}[13pt][7pt]{not given} \\

\raisebox{0pt}[13pt][7pt]{$\Upsilon \to \mu^+ \tau^-$} & 
\raisebox{0pt}[13pt][7pt]{negligible} & 
\raisebox{0pt}[13pt][7pt]{$2 \times 10^{-6}$} & 
 
\raisebox{0pt}[13pt][7pt]{not given} \\

\raisebox{0pt}[13pt][7pt]{$\mu^- \to e^- \gamma$} & 
\raisebox{0pt}[13pt][7pt]{?} & 
\raisebox{0pt}[13pt][7pt]{0} & 
 
\raisebox{0pt}[13pt][7pt]{$4.9 \times 10^{-11}$} \\

\raisebox{0pt}[13pt][7pt]{$\mu^- \to e^- e^+ e^-$} & 
\raisebox{0pt}[13pt][7pt]{?} & 
\raisebox{0pt}[13pt][7pt]{0} & 
 
\raisebox{0pt}[13pt][7pt]{$1.0 \times 10^{-12}$} \\\hline
\end{tabular}
\end{center}
\vspace*{-5mm}
\end{table}

With significant exceptions, SM
(with FSD) estimates are all far below present experimental bounds and
are hence not so interesting.  An exception is the process $\mu^- \to
e^-e^+e^-$, where one could get a branching ratio of $10^{-3}$
(limit $10^{-12}$), if one applied FSD naively.  

The parameter-free calculations in DSM give branching ratios which
are in general larger but still below present experimental limits.
It is important to note that because
of SSD the branching ratios of transmutational leptonic decays are
automatically zero to first order (Figure \ref{masmat}).  
The $\pi^0$ decay is of particular
interest as being less than one order from the experimental bound.

\section{Photo-transmutation}
We have studied\cite{photrans} the following process, mainly for DSM:
\begin{equation}
\gamma +\ell_\alpha  \longrightarrow \gamma +\ell_\beta , \quad \alpha
\ne \beta.
\end{equation}

There are two points to note.  First, the kinematics is unfamiliar, with 
the mass eigenstates
being varying linear combinations of the $\tau,\mu,e$
states.  
The other point is that standard formulae
for summing $\gamma$ matrix traces cannot be applied directly.  We
calculated the individual spin and polarization amplitudes and then
summed them by hand.

\begin{figure}
\centering
\epsfxsize120pt
\includegraphics[angle=-90,scale=0.3]{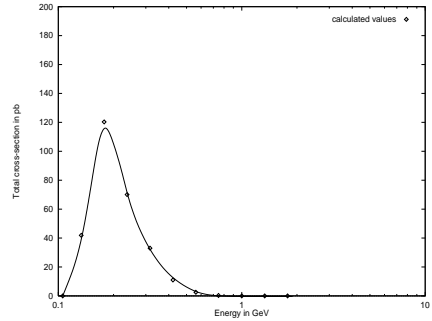}
\caption{Total cross section for $\gamma e \to \gamma \mu$.}
\label{total}
\vspace*{-7mm}
\end{figure}

We calculated the cross sections for: 
$\gamma e \to \gamma \mu,\ \gamma e \to \gamma \tau,\ \gamma \mu \to
\gamma \tau$, leaving out $\tau$-initiated reactions as 
experimentally unrealistic at present.  
\begin{figure}
\centering
\epsfxsize120pt
\includegraphics[angle=-90,scale=0.4]{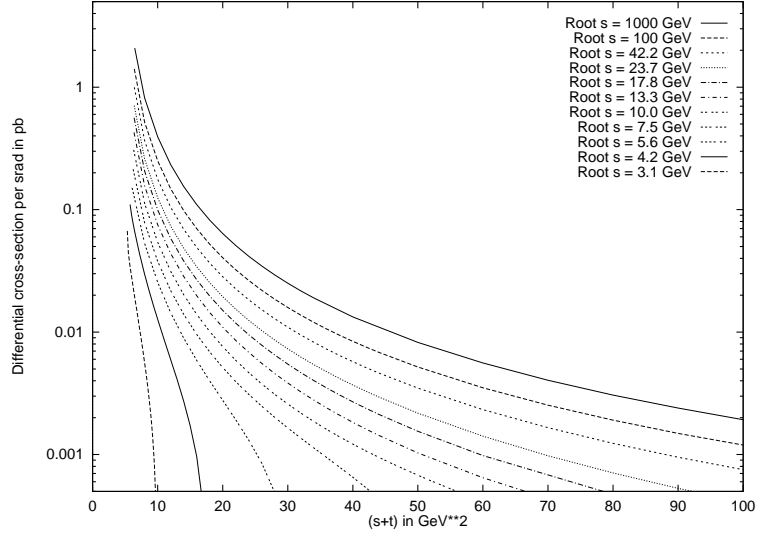}
\caption{Differential cross sections for $\gamma e \to \gamma \tau$.}
\label{cross}
\vspace*{-5mm}
\end{figure}
A sample of the DSM results is
presented in Figure \ref{cross}.
We get in general:
$$\gamma \mu \to \gamma \tau > \gamma e \to \gamma \tau > \gamma e \to
\gamma \mu.$$
However, at low energies $\gamma e \to
\gamma \mu$ becomes quite sizeable, Figure \ref{total},
with the total cross section having a peak of $\sim$ 100 pb at c.m.\
energy $\sim$ 200 MeV.

Calculations in SM depend on further assumptions.  If the Dirac masses
of the neutrinos are hierarchical, then above $\tau$ the
cross sections are approximately given by scaling
those of DSM with the relevant rotation matrix elements.  For example
$\gamma \mu \to \gamma \tau$ at $\sqrt{s}= 17.8$ GeV is $2-3$
orders smaller.

\section{Possible experimental tests in the near future}
Since only the DSM estimates and calculations are parameter-free,
while the SM results presented here are subject to further assumptions
and uncertainties, we shall point out tests for DSM only here.

The estimates for the transmutational decay
modes: $\pi^0
\to \mu^- e^+,\ \psi \to \tau^- \mu^+,\ \Upsilon \to \tau^- \mu^+$
could be near experimental limits and sensitivites, for LEP, BEPC
and B-factories. 

For photo-transmutations, one may consider virtual 
$\gamma$ from $e^+
e^-$ colliders.  Above $\tau,\ e \to \tau$ is more important,
while below it is $e \to \mu$.  Again, LEP
and/or BEPC may provide tests\cite{study}.

\section{Conclusions}
\hspace*{\parindent}$\bullet$ Fermion transmutation
necessarily occur
in both SM and DSM.\\
\hspace*{\parindent}$\bullet$ The SM results are in general smaller 
than DSM, with uncertainties.\\
\hspace*{\parindent}$\bullet$ The DSM calculations are entirely 
parameter-free.    There are no
violations of data in all the cases we were able to consider.\\
\hspace*{\parindent}$\bullet$ Experimental tests of DSM predictions 
seem feasible in
the near future: (1) decays of $\pi^0,\psi,\Upsilon$,
(2) photo-transmutation of $\gamma e \to \gamma 
\tau$ at high and $\gamma e \to \gamma \mu$ at low energy,
(3) other processes e.g.\ $e^+ e^- \to e^+ \mu^-$.\\
\hspace*{\parindent}$\bullet$ Transmutation leads to
exciting new physics that has to be explored.


\end{document}